# Enhanced light emission from plasmonic nanostructures by molecules

Yuqing Cheng,[1] Jingyi Zhao,[1] Te Wen,[1] Guantao Li,[1] Jianning Xu,[1] Qihuang Gong,[1,2] and Guowei Lu[1,2,*]

[1]*State Key Laboratory for Mesoscopic Physics & Collaborative Innovation Center of Quantum Matter, Department of Physics, Peking University, Beijing 100871, China*
[2]*Collaborative Innovation Center of Extreme Optics, Shanxi University, Taiyuan, Shanxi 030006, China*
[*] *e-mail address: guowei.lu@pku.edu.cn*

Interaction between plasmonic nanostructures and molecules is modeled based on the concept of quantized optical cavity for surface enhanced Raman scattering process. We have found that the background emission from plasmonic nanostructures is not constant as speculated ordinarily, it is enhanced accompanying with the molecules Raman scattering. The plasmonic nanostructures not only scatter elastically the energy coupling from the molecules excited states, but also radiate it inelastically as surface plasmon emission partly resulting an enhanced background. According to single nanoparticle experiments, the model reveals that the background fluctuations is mainly due to the induced field of the molecules, which increases the local field felt by the nanostructures that was often overlooked in the past. These findings suggest considering the plasmonic nanostructures and molecules as a hybrid entity to analyze and optimize the surface enhanced spectroscopy.

Surface enhanced Raman scattering (SERS) has been widely investigated for about forty years due to exciting opportunities for applications of vibrational spectroscopy into a structurally sensitive molecule and nanoscale probe. A broad consensus in SERS community is that the SERS mechanism is mainly due to the local electromagnetic enhancement of localized surface plasmon resonances (LSPs) supported in metallic nanostructures.[1-5] While a broad continuum emission called "background" is always observed in SERS spectra. Such accompanying background has often been ignored by using background subtraction methods in experiments since the SERS background was supposed to be a stable continuum. Much of this discussion of the SERS background was confined to the initial years after the discovery of SERS, but its understanding and mechanism are not reported always.[6,7] Nowadays a solid consensus is that plasmonic nanostructures can emit light at their LSPs band under excitation of electrons or photons, although the mechanism is still the subject of much debate. For instance, photoluminescence (PL) from metal thin films under light excitation was first reported in 1969.[8] Later, it was found that roughened surfaces increase the PL efficiencies through the enhancement of local optical fields.[9] Recently, the PL from metallic nanostructures has been observed with considerably higher efficiencies.[10-12] The SERS continuum background can be correlated with the PL of the metal nanostructures.

It is well known that the PL or scattering intensity of plasmonic nanostructures is non-bleaching and non-blinking. Previously, the influences of adsorbed molecules on the localized field was presumed to be negligible, then the affection of the molecules on light emission of the metallic nanostructures was supposed to be very weak. Hence, the SERS background was often supposed to be a stable continuum. While some studies showed that the SERS broad background fluctuated with SERS vibrational Raman peaks.[13] In other words, the light emission from the nano-metal is changeable during SERS due to the interaction between the nano-metal and the molecules. However, the SERS background fluctuations were overlooked usually in the conventional electromagnetic theory.[14,15]

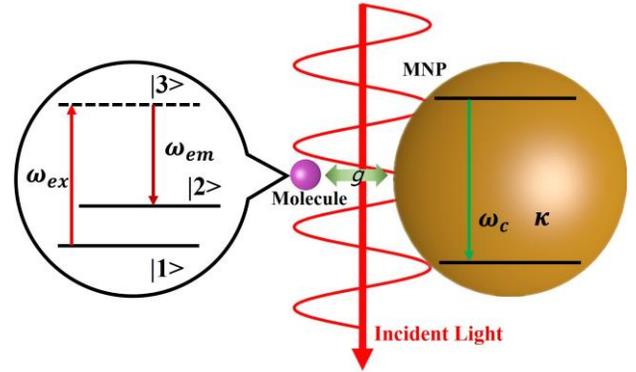

FIG. 1. (Color online) Schematic diagram of the model describing the interaction between MNP and molecule which are excited by the incident electromagnetic field with frequency $\omega_{ex}$. The coupling coefficient is $g$. The energy levels picture presents the Stokes Raman process.

In this study, we develop a phenomenological theory model based on the concept of quantized optical cavity to simulate the interaction of the molecules and the nearby plasmonic nanostructures for the SERS process. The plasmonic nanostructure was modeled as an optical nano-resonator. The nano-resonator not only scatter the incident light elastically at the excitation wavelength but also emit light through the plasmons decay at its LSP band inelastically.[16] The Raman scattering process of the molecules was modeled as a three-level atom system in the Λ configuration whose excited state was considered to be overlay with the nano-resonator LSP band.[17] It was found that the molecule enables an enhanced plasmon emission from the plasmonic nano-resonator. Furthermore, the experiment of the SERS blinking based on a single gold nano-flower (GNF) was performed to demonstrate the fluctuations of SERS background. The theoretical and experimental results agree qualitatively, i.e. the SERS background is inevitable and enhanced accompanying always with molecules' SERS vibrational Raman peaks. Both the coupling effect and the induced field of the molecule contribute to the light emission enhancement of the GNF.

Fig.1 shows the schematic diagram of the model. The metal nanoparticle (MNP) is considered as a nano-cavity with a resonance frequency $\omega_c$ and a total decay rate $\kappa$. The molecule is considered as a three-level atom with the vibrational energy level |1> and |2>, and virtual level |3>. For simplicity, we concentrate on the Stokes Raman emission as an example to demonstrate the interaction between the molecule and the nano-cavity. A quasi-continues-wave laser beam with frequency $\omega_{ex}$ couples the ground state |1> and the virtual level state |3>, where the energy difference between |1> and |3> is $\omega_{31} = \Omega_3 - \Omega_1$. We assume that $\omega_{31} = \omega_{ex}$. Latter, the excited electron at |3> jumps into |2> and emits a photon with frequency $\omega_{em} = \omega_{32} = \Omega_3 - \Omega_2$. Apparently, $\omega_{32} = \omega_{31} - \omega_{21} = \omega_{ex} - \omega_{21}$ or $\Delta\omega = \omega_{ex} - \omega_{em} = \omega_{21}$. Now, we obtain the frequency relation between the excitation photon and the emitted photon. Regarding to the MNP plasmonic resonator, the free energy of the LSP mode $a$ with the resonance frequency $\omega_c$, and Hamiltonian is described as $H_c = \omega_c a^\dagger a$. For the free atom, Hamiltonian is written as $H_m = \sum_{j=1}^{3}\Omega_j \sigma_{jj}$, where $\sigma_{ij}$ (i, j=1,2,3) are operators in the atom subspace. Specifically, $\sigma_{ij}$ ($i \neq j$) are the transition operators from state |j> to state |i>. Therefore, the free Hamiltonian of the atom-cavity system without any interaction is written as:

$$H_0 = H_c + H_m = \omega_c a^\dagger a + \sum_{j=1}^{3}\Omega_j \sigma_{jj} \quad (1)$$

In addition, the interaction Hamiltonian is described as

$$H_I = g(\sigma_{23}a^\dagger + a\sigma_{32}) + \mu_c(a^\dagger E_1 e^{-i\omega_{ex}t} + aE_1 e^{i\omega_{ex}t}) + \mu_{13}(\sigma_{31}E_2 e^{-i\omega_{ex}t} + \sigma_{13}E_2 e^{i\omega_{ex}t}) \quad (2)$$

where the first term describes that the LSP mode is coupled with states |3> and |2>, and $g$ is the coupling constant. The second and the third term describe the processes that the excitation field is coupled with LSP mode and states |1> and |3>, respectively, and $\mu_c$ and $\mu_{13}$ are the respective coupling constants. $E_1$ and $E_2$ are the respective localized field amplitudes that the MNP and the atom feel. Hence, the Hamiltonian of this system is given by

$$H = H_0 + H_I \quad (3)$$

The dynamics of these modes can be solved by the equations:

$$\dot{a} = i[H,a] - \kappa a = (-i\omega_c - \kappa)a - ig\sigma_{23} - i\mu_c E_1 e^{-i\omega_{ex}t} \quad (4)$$

$$\dot{\sigma}_{23} = i[H,\sigma_{23}] - \gamma\sigma_{23} = (-i\omega_{32} - \gamma)\sigma_{23} + iga\sigma_{z32} - i\mu_{13}E_2\sigma_{21}e^{-i\omega_{ex}t} \quad (5)$$

Where $\kappa$ and $\gamma$ are the total decay of the MNP and the atom, respectively and $\sigma_{z32} = \sigma_{33} - \sigma_{22}$. The equations can be solved rigorously, and the formal of the solutions are:

$$a = A_1 e^{(-i\omega_1-\kappa_1)t} + A_2 e^{(-i\omega_2-\kappa_2)t} + A_3 e^{-i\omega_{ex}t} \quad (6)$$

$$\sigma_{23} = B_1 e^{(-i\omega_1-\kappa_1)t} + B_2 e^{(-i\omega_2-\kappa_2)t} + B_3 e^{-i\omega_{ex}t} \quad (7)$$

where $\omega_1$, $\omega_2$ and $\kappa_1$, $\kappa_2$ are given as below: $\omega_1, \omega_2 = -\frac{1}{2}Im(D_1 \mp D_2)$, and $\kappa_1, \kappa_2 = -\frac{1}{2}Re(D_1 \mp D_2)$, where we define $D_1 = -\kappa - \gamma - i\omega_c - i\omega_{32}$ and $D_2 = \sqrt{[(\kappa-\gamma)+i(\omega_c-\omega_{32})]^2 - 4g^2}$. The coefficients $A_1$ and $A_2$ are complex amplitudes of modes $\omega_1$, $\omega_2$ for particle's operator $a$. $B_1$ and $B_2$ are complex amplitudes of modes $\omega_1$, $\omega_2$ for molecule's operator $\sigma_{23}$. $A_3$ and $B_3$ are complex amplitudes of scattering modes for $a$ and $\sigma_{23}$, respectively.

For weak coupling system, we assume that $g \ll \kappa - \gamma$, thus the eigen frequencies turn into $\omega_1, \omega_2 \approx \omega_c, \omega_{32}$, and the decay rates turn into $\kappa_1, \kappa_2 \approx \kappa, \gamma$. In the case of weak coupling, we ignore all the $g^2$ items and obtain these coefficients:

$$A_1 = \frac{iE_1\mu_c}{i(\omega_c-\omega_{ex})+\kappa} + \frac{gE_2\mu_{13}\sigma_{21}}{[i(\omega_c-\omega_{ex})+\kappa][i(\omega_c-\omega_{em})+\kappa-\gamma]};$$

$$A_2 = -\frac{gE_2\mu_{13}\sigma_{21}}{[i(\omega_{em}-\omega_{ex})+\gamma][i(\omega_c-\omega_{em})+\kappa-\gamma]};$$

$$B_1 = \frac{gE_1\mu_c\sigma_{z32}}{[i(\omega_c-\omega_{ex})+\kappa][i(\omega_c-\omega_{em})+\kappa-\gamma]};$$

$$B_2 = -\frac{iE_2\mu_{13}\sigma_{21}}{i(\omega_{em}-\omega_{ex})+\gamma} - \frac{gE_1\mu_c\sigma_{z32}}{[i(\omega_{em}-\omega_{ex})+\gamma][i(\omega_c-\omega_{em})+\kappa-\gamma]};$$

Using the input-output relation $\langle a_{out}\rangle = \sqrt{2\kappa_{exa}}\langle a\rangle$, where $\langle \hat{Q}\rangle$ is the quantum average of the operator $\hat{Q}$ and $\kappa_{exa}$ is the outgoing coupling rate of the MNP, the detected intensity of light emission from the nano-cavity can be evaluated by

$$I_{full-a}(\omega) = \Re\left[\int_0^\infty \langle a_{out}^\dagger(\tau+t)a_{out}(t)\rangle e^{-i\omega\tau}d\tau\right]$$

$$= \Re\left[\int_0^\infty \left[\frac{1}{T}\int_0^T a_{out}^\dagger(\tau+t)a_{out}(t)dt\right]e^{-i\omega\tau}d\tau\right] \quad (8)$$

Where $\Re[Q]$ stands for the real part of Q. $I_{full-a}(\omega)$ includes PL $I_{PL-a}(\omega)$ and scattering $I_{SC-a}(\omega)$. After filtering the input laser field and using the quantum regression theorem, we obtain the PL intensity of the nano-cavity from $I_{full-a}(\omega)$ as

$$I_{PL-a}(\omega) = 2\kappa_{exa}\left[|A_1|^2\left(\frac{1-e^{-2\kappa_1 T}}{2\kappa_1 T}\right)\frac{\kappa_1}{(\omega-\omega_1)^2+\kappa_1^2} + |A_2|^2\left(\frac{1-e^{-2\kappa_2 T}}{2\kappa_2 T}\right)\frac{\kappa_2}{(\omega-\omega_2)^2+\kappa_2^2}\right] \quad (9)$$

where $T$ is effective interaction time between the electrons and light wave packet. Similarly, by using the relation $\langle \sigma_{out}\rangle = \sqrt{2\kappa_{exb}}\langle\sigma_{23}\rangle$, where $\kappa_{exb}$ is the outgoing coupling rate of the atom, the detected intensity of light emission from the atom can be evaluated by

$$I_{full-\sigma}(\omega) = \Re\left[\int_0^\infty \langle \sigma_{out}^\dagger(\tau+t)\sigma_{out}(t)\rangle e^{-i\omega\tau}d\tau\right]$$

$$= \Re\left[\int_0^\infty \left[\frac{1}{T}\int_0^T \sigma_{out}^\dagger(\tau+t)\sigma_{out}(t)dt\right]e^{-i\omega\tau}d\tau\right] \quad (10)$$

then the PL intensity of the atom from $I_{full-\sigma}(\omega)$ as

$$I_{PL-\sigma}(\omega) = 2\kappa_{exb}\left[|B_1|^2\left(\frac{1-e^{-2\kappa_1 T}}{2\kappa_1 T}\right)\frac{\kappa_1}{(\omega-\omega_1)^2+\kappa_1^2} + |B_2|^2\left(\frac{1-e^{-2\kappa_2 T}}{2\kappa_2 T}\right)\frac{\kappa_2}{(\omega-\omega_2)^2+\kappa_2^2}\right] \quad (11)$$

Hence, total detected emission spectrum would be

$$I_{PL}(\omega) = I_{PL-a}(\omega) + I_{PL-\sigma}(\omega) \quad (12)$$

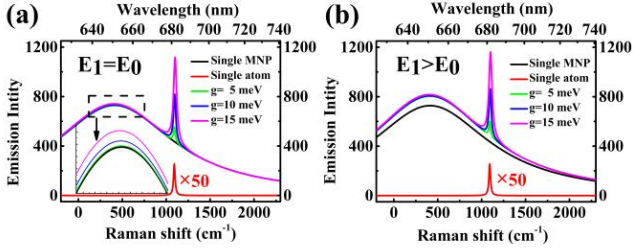

FIG. 2. (Color online) Simulated spectra of single MNP (black), single atom (red) and coupled system (green, blue and purple) for different coupling coefficients $g$. The inset of (a) presents the enhanced background by enlarging. $E_1$ is set as $E_1 = E_0$ for (a) and $E_1 > E_0$ for (b).

So far, we deduce the light emission spectra from both the MNP and the atom with the coupling coefficient $g$. For an extreme case ($g=0$), which suggest that there is no interaction between MNP and the atom and the coefficients would be given in such form: $A_1 = \frac{iE_1\mu_c}{i(\omega_c-\omega_{ex})+\kappa}$, $A_2 = 0$, $B_1 = 0$, $B_2 = -\frac{iE_2\mu_{13}\sigma_{21}}{i(\omega_{em}-\omega_{ex})+\gamma}$. The modes and decays are regressed to the original form: $\omega_1, \omega_2 = \omega_c, \omega_{32}, \kappa_1, \kappa_2 = \kappa, \gamma$. Thus the PL spectrum from each operator ($a, \sigma_{23}$) presents a single Lorentz line shape:

$$I_{PL-a}(\omega)_{g=0} = 2\kappa_{exa}|A_1|^2\left(\frac{1-e^{-2\kappa_1 T}}{2\kappa_1 T}\right)\frac{\kappa_1}{(\omega-\omega_1)^2+\kappa_1^2} \quad (13)$$

$$I_{PL-\sigma}(\omega)_{g=0} = 2\kappa_{exb}|B_2|^2\left(\frac{1-e^{-2\kappa_2 T}}{2\kappa_2 T}\right)\frac{\kappa_2}{(\omega-\omega_2)^2+\kappa_2^2} \quad (14)$$

which is corresponding to the PL from the single emitter (MNP or atom, $E_1 = E_2 = E_0$). Representative simulation spectra from eqn.12-14 are plotted in Fig.2 based on conditions that give $\mu_c E_0 = 1\ eV$, $\mu_{13}E_0 = 0.08\ eV$, $\omega_{ex} = 1.964\ eV$, $\omega_c = 1.913\ eV$, $\omega_{em} = 1.828\ eV$, $\kappa = 0.103\ eV$, $\gamma = 1.344\ meV$, $\kappa_{exa} = 1\ eV$, $\kappa_{exb} = 0.08\ eV$. Black and red curves are free single MNP and atom emission spectra interacted with the incident light only, respectively (according to eqn. 13 and 14). We obtain that the emission spectrum shows a Lorentz line as narrow Raman line. The background is light emission by LSP mode radiative decay of the nano-resonator.

For $g \neq 0$, i.e. there is interaction between the MNP and the atom. As seen from Fig.2, and the light emission from the molecule and plasmonic nanostructure both increase simultaneously. The coefficient $A_1$ for operator $a$ is the key parameter related with the SERS background emission. We define the background enhancement factor for the nano-resonator:

$$M_1 = \left|\frac{A_1(g\neq 0)}{A_1(g=0)}\right|^2 + \frac{\kappa_{exb}}{\kappa_{exa}}\left|\frac{B_1(g\neq 0)}{A_1(g=0)}\right|^2$$

$$= \left|\frac{E_1}{E_0} + \frac{gE_2\mu_{13}\sigma_{21}}{iE_0\mu_c[i(\omega_c-\omega_{em})+\kappa-\gamma]}\right|^2 + \frac{\kappa_{exb}}{\kappa_{exa}}\left|\frac{E_1}{E_0}\frac{g\sigma_{z32}}{i(\omega_c-\omega_{em})+\kappa-\gamma}\right|^2.$$

We assume that $\langle\sigma_{21}\rangle = 0$, $|\langle\sigma_{21}\rangle|^2 = 0.25$ (Supplemental Materials). It can be easily seen that $M_1 > 1$. Meanwhile, we can also define the Raman enhancement factor for the molecule:

$$M_2 = \frac{\kappa_{exa}}{\kappa_{exb}}\left|\frac{A_2(g\neq 0)}{B_2(g=0)}\right|^2 + \left|\frac{B_2(g\neq 0)}{B_2(g=0)}\right|^2$$

$$= \left|\frac{E_2}{E_0} + \frac{gE_1\mu_c\sigma_{z32}}{iE_2\mu_{13}\sigma_{21}[i(\omega_c-\omega_{em})+\kappa-\gamma]}\right|^2 + \frac{\kappa_{exa}}{\kappa_{exb}}\left|\frac{E_2}{E_0}\frac{g}{i(\omega_c-\omega_{em})+\kappa-\gamma}\right|^2.$$

Green, blue and purple curves in Fig.2 are the emission spectra by the coupled system (eqn.12) with different coupling coefficients $g$. Each coefficient, $g$ corresponds to the near field enhancement which the atom feels, that is, $g = 5\ meV$, $10\ meV$, $15 meV$ correspond to $E_2 = 5E_0$, $E_2 = 10E_0$, $E_2 = 15E_0$. For these curves, the electromagnetic field, $E_1$ that the MNP feels is set as $E_1 = E_0$ in Fig.2a. Here, we assumed at first that the induced field of the molecule does not influence the field felt by the MNP. While $E_2$ that the atom feels enhancement a lot due to the LSP near field effect of the MNP. Besides, the molecule Raman enhancement factor $M_2$ are calculated given as $M_2 = 27$ (green), 113 (blue), 275 (purple). It should be noted that for smaller value of $\mu_{13}$, the factor $M_2$ becomes larger that would be much closer to actual SERS enhancement due to the coupling effect.

The increase of $M_1$ is due to the interaction between the MNP and the molecule. When the coupling coefficient, $g$ becomes larger, the factor $M_1$ is larger, and the same as the Raman intensity of the molecule. While we obtain a slight (less than 2%) enhancement for factor $M_1$ (Fig.2a). That is not enough to explain the intensity change in experiment shown in Fig.3. Then, we tuned $E_1$ and set as $E_1 = 1.15E_0$ and we obtain the considerable enhancement of the background emission as shown in Fig.2b. The intensity of the background is proportional to $|E_1|^2$, so the domain parameter that influences $M_1$ is $E_1$, but not the coupling coefficient $g$. That implies that the induced field produced by the polarization of the molecule cannot be overlooked again. Regarding to the local field $E_1$ felt by the MNP, in the research field of strong coupling, it is acceptable that the induced field of large size quantum dot become comparable or higher than the external excitation field.[18] In the present SERS system, giant localized electromagnetic field induced by "hot spot" would polarizes the molecule and then produces an induce field to interact with the MNP in turn. Supposed the localized field enhancement reaching ~100-fold and the effective size of "hot spot" less than 5 nm, then the induced field of the molecules becomes detectable and the assumption of the field $E_1$ by the MNP is reasonable (Supplemental Materials).

To demonstrate the plasmonic nanostructure and the molecule enhance mutually their light emission in SERS process, we employ single gold nano-flower (GNF) to perform single nanoparticle based SERS blinking. The advantage of the GNFs is that the "hot spots" are available easily so that the Raman blinking happened more often. Fig.3a shows the schematic diagram of our system in experiment and the Raman spectra blinking. We speculate that the molecule migration or rotation at the hot spot site leads to the SERS spectra temporal blinking, i.e. converting between the states of $g = 0$ and $g \neq 0$.[19-22] In other words, when the spectrum is at the background level (without any molecule Raman signal), the coupling coefficient can be assumed as $g \approx 0$, the background spectrum is attributed to the PL of the GNF. When the spectrum burst presenting high molecule Raman peak signal, the coupling coefficient could be assumed as $g \neq 0$, which shows the light emission spectrum of the coupled system.

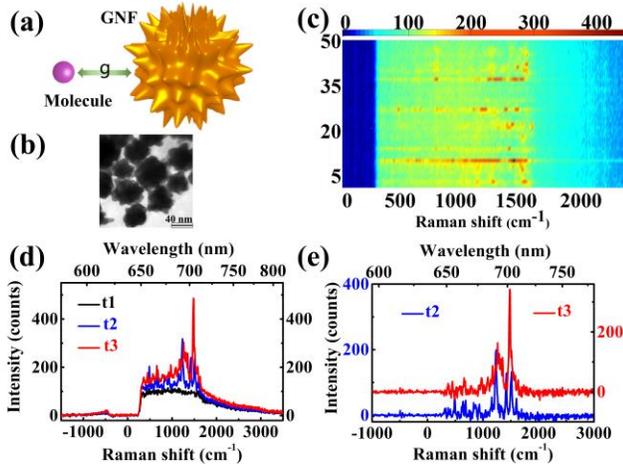

FIG. 3. (Color online) (a) Schematic diagram of SERS system. (b) Representative SEM image of the GNFs with the scale bar of 70 nm. (c) SERS blinking of the GNF and 4-mercaptopyridine molecules system. (d) SERS spectra selected from Fig.3c. (e) Difference from (d), blue and red line stands for $I_{t2} - 1.24 \times I_{t1}$ and $I_{t3} - 1.53 \times I_{t1}$ separately.

A microspectroscopy system based on an inverted optical microscope was developed to combine multifunctional optical measurements. The setup allows us to identify single isolated nanoparticles and *in situ* measure the luminescence and Raman spectra near a single nanoparticle. The experimental configurations were described in detail in the previous study.[23] In the measurements, we used an oil-immersion objective lens. A continue wave laser at wavelength of 633 nm was used as the excitation light with excitation power of ~160 μW. The GNFs used in our experiments with a diameter of 70 ± 10 nm were synthesized by the wet chemical method. The nanoparticles were immobilized onto silane functionalized glass coverslips with an average interparticle spacing of several micrometers. After that the chip with immobilized GNFs was treated with 1 μM 4-mercaptopyridine molecules solution for over 3 hours and dried in $N_2$ gas flow. Representative 50 consecutive spectra recorded from a single GNF with acquirement time of 0.1 s are plotted in Fig.3c. We observed Raman blinking phenomena, and both the background due to the GNF and the Raman peaks of the molecule are fluctuation. Three representative spectra were selected for comparison as shown in Fig.3d. According to above conclusion in Fig. 2, when $g$ is larger, the blinking Raman peaks goes higher. Therefore, the red and blue curves in Fig.3d stand for the strong coupling and the black line stands for the weakest coupling ($g \approx 0$).

As previously stated, the SERS background results from the LSP mode $a$ and presents the Lorentz line shape the same as the uncoupled LSP mode. Actually, by correlating scattering, PL and SRES spectra in previous experiments, the SERS background ascribes to the PL from the plasmonic nanostructure.[24,25] Hence, we can assume that the black line stands for the PL from the GNF approximately. We can obtain a Raman spectrum without broad background by subtracting the PL spectrum of the GNF multiplied by an enhancement factor from the SERS blinking spectra. Fig.3e shows the resulted curves from the spectra in Fig.3d. Corresponding multiple for background enhancement factor are 1.24 and 1.53 for blue and red curves separately. That means the induced field of the molecule increase the effective local field intensity $E_1$ felt by the GNF ~11% and 24%, respectively. And the subtracted spectra do not present the background, which implies that the PL from the GNF mainly contribute the SERS background. Although the origin or fluctuation of SERS background could be ascribed to other factor like carbon contaminate etc., our calculations and experiment results agree qualitatively, and provide a self-consistent understanding.

In conclusion, we have investigated the fluctuations of background continuum in the SERS process based on the concept of quantized optical cavity. We have found that the inevitable presence of background resulted from the metallic nanostructure plasmon emission is not stable as speculated ordinarily, which increased simultaneously when the molecules' Raman scattering was enhanced. The plasmonic nanostructures can not only scatter the coupling energy from the excitation states of the molecules directly, but also convert it into surface plasmon that decays radiative partly. The model enables us to understand qualitatively the background fluctuations of the SERS blinking spectra based on single GNF in experiment. The model reveals that the background fluctuations in experiment were mainly due to the induced field of the molecules, which increase the local field felt by the nanostructures. The SERS background is changeable and it can be another indicator for the interaction strength between the plasmonic nanostructures and the molecule. These findings suggest considering the SERS as an entity system to analyze and optimize the interaction signal. The concept would also be effective for other surface enhanced spectroscopy such surface enhanced fluorescence.[26]

## ACKNOWLEDGEMENTS

This work was supported by the National Key Basic Research Program of China (grant no. 2013CB328703) and the National Natural Science Foundation of China (grant nos. 61422502, 11374026, 61521004, 11527901).